# Gendered behavior as a disadvantage in open source software development

*Balazs Vedres, Orsolya Vasarhelyi*
*Central European University, Department of Network and Data Science*

**Abstract**

Women are severely marginalized in software development, especially in open source. In this article we argue that disadvantage is more due to gendered behavior than to categorical discrimination: women are at a disadvantage because of what they do, rather than because of who they are. Using data on entire careers of users from GitHub.com, we develop a measure to capture the gendered pattern of behavior: We use a random forest prediction of being female (as opposed to being male) by behavioral choices in the level of activity, specialization in programming languages, and choice of partners. We test differences in success and survival along both categorical gender and the gendered pattern of behavior. We find that 84.5% of women's disadvantage (compared to men) in success and 34.8% of their disadvantage in survival are due to the female pattern of their behavior. Men are also disadvantaged along their interquartile range of the female pattern of their behavior, and users who don't reveal their gender suffer an even more drastic disadvantage in survival probability. Moreover, we do not see evidence for any reduction of these inequalities in time. Our findings are robust to noise in gender recognition, and to taking into account particular programming languages, or decision tree classes of gendered behavior. Our results suggest that fighting categorical gender discrimination will have a limited impact on gender inequalities in open source software development, and that gender hiding is not a viable strategy for women.

**Keywords**

gender inequality, gendered behavior, software development, open source

**Significance:**

Software development is vital in shaping society, from finance to social relationships, and women are largely excluded. Understanding the relative importance of categorical gender and the gendered pattern of behavior in the marginalization of women can fundamentally redirect policy initiatives. Should policy target those who discriminate against women, or should it rather target choices that become available or unavailable to men and women during their careers? We found that 84.5% of women's disadvantage in success, and 34.8% of their disadvantage in survival, is due to gendered behavior. Policies targeting the gender category, or individual strategies of hiding one's gender, are ineffective against discrimination by the pattern of behavior. How can policy intervene in the use of such behavioral discrimination?

**Introduction**

Women suffer a considerable disadvantage in information technology: their proportion in the workforce is decreasing, and they are especially underrepresented in open source software development. The proportion of women in computing occupations has been steadily declining from 36% in 1991 to 25% today (1–3). In open source software only about 5% of the developers are women (4). Women receive less recognition for their work, and they exit their computing occupation careers with higher probability. Women suffer from a gender wage gap in STEM – and especially in computer programming – more so than in other fields (5): they suffer a net wage disadvantage of 7% that has not decreased over the past two decades (6). Many women quit their computing occupation careers in the middle (7). These developments are puzzling, especially in the face of a favorable shift in public consciousness, and considerable private and public policy efforts to counter gender discrimination. With accumulating evidence of the benefits of gender diversity in teams (8–10), it is clear that marginalization of women in software development leads to major societal costs.

In this article we analyze a large dataset of open source software developers to answer the question: are women at a disadvantage because of who they are, or because of what they do? Typically, gender discrimination is conceptualized as categorical discrimination against women (11); however, as much of the scholarship in gender studies had shown, to understand gender inequalities one needs to shift the focus to the gendered pattern of behavior (12, 13): The more likely causes of discrimination are actions that are typical of men and women, rather than the gender category of the person (14–16). Women in leadership roles often feel compelled to (or are expected to) follow male behavioral traits (17), just as men in feminine occupations take on female-like behavioral traits (18), and the choice of collaborators and mentors often follows gender homophily (19).

While categorical gender discrimination is an easy target for policies, discrimination based on behavioral expectations are more difficult to counter. Recently Google was sued by women for categorizing women as 'front-end' developers without reason, blocking their access to higher pay and faster promotion that 'back-end' developers enjoy, who are more likely to be male (20). This also underscores that when we analyze the gendered pattern of behavior, we should not assume that such behavior is a result of free choice. In fact, the history of computing occupations is also a history of marginalizing women from an increasing number of specializations (21). Thus far there have been no analysis based on large data in a contemporary setting, to analyze behavioral traces, and to assess the relative weight of categorical and behavioral gender in gender inequality.

Our data source is GitHub: the most popular online open source software project management system, which provides an opportunity to track the behavior of software developers directly, identify gender from user names, and observe success and survival (22, 23). In open source software development the most important payoff to participants is reputation (24), hence we operationalize success as the number of users declaring interest in ones work by "starring" a repository. As a second dependent variable we analyze differences in the odds of sustaining open source development activity over a one year period subsequent to our data collection time window.

Using data about behavior in a large sample allows us to construct a measure of femaleness of observed behavioral choices over the entire career, as a measure of gender typicality. This approach has a long history, using survey data (12, 25, 26), and more recently with behavioral trace data in diverse settings (27–29). In addition to the interval scale gendered behavioral dimension, we also identify multiple kinds of gendered behavioral patterns using a decision tree classification approach, and we assess the relative explanatory power of one behavioral dimension when controlling for multiple patterns of behavior.

We first compare men and women: users who display a recognizable gender on their profile, but we also analyze data of users with un-identifiable gender. The first question is whether gendered behavior makes any difference at all, or is it only the gender category, that relates to female disadvantage. If gendered behavior is related to outcomes, is that relationship the same for both women and men? Are there signs of change in patterns of gendered disadvantage?

It is also important to analyze gendered behavior of those who do not readily reveal their gender. Scholars have discussed the potential of online collaborations to mitigate gender inequalities, as it is easier to manipulate or hide gender identity online, compared to face-to-face settings (30–32). Our first question here is whether we see evidence for surrounding users recognizing the gender from the behavior of focal users that are hiding their categorical gender. Our second question is whether success and survival for unknown-gender users are related to their gendered behavior as well.

**Empirical Setting and Data**

Github (github.com) is a social coding platform that allows software engineers to develop and publish software together, recording their contributions to a collaborative activity. It is the most popular web-based 'git' software repository hosting and version tracking service, with 20 million users and over 57 million private and public repositories in May, 2018. Working in repositories collaboratively can lead to success through visibility and reputation, which helps developers to be noticed by potential employers (24, 33, 34) We used coding and collaboration activity to conceptualize individual careers.

The empirical basis of this study is a data set acquired via githubarchive.org between 2009-02-19 and 2016-10-21 about the following: creation of a repository, push to a repository, opening, closing and merging a pull request. To collect information about users' names, e-mail addresses, number of followers, number of public repositories and the date they joined GitHub, we sent calls to the official Github users API.

Since users do not list their gender directly, we infer each person's gender using their first names. This is a commonly and successfully used method in Western societies (35). In this work, we relied on the 2016 US baby name dataset published by the US Social Security Administration annually. (SSA 2016). GitHub makes it optional to add full names to profiles; therefore we infer first names from emails as well. Due to some names being used for both males and females, we assign a probability of being male to each candidate based on the fraction of times their first name was assigned to a male baby in the name dataset. We define gender probability cutoffs of 0.1 and 0.9 consistent with previous studies (34). Our gender recognition yielded 11.87% females and 88.13% males out of all users with names. All in all we found 194,010 females, 1,441,130 males, and 6,163,370 unknowns. See SI, and S1 and S2 specifically for details of our gender recognition workflow and results.

We decided to filter users by their level of activity, as there are many users who establish a GitHub account with hardly any subsequent developer engagement (but use GitHub, for example, as a web hosting platform). First we selected those 1,634,373 users in our data set with at least 10 traces of activity over their careers. Then we deleted 1,604 users for evidence of being artificial agents (having a substring, like "bot", "test", "daemon", "svn2github", "gitter-badger" in their usernames). As we were interested in patterns of gendered behavior (for which we encountered resource and time intensive data crawling challenges regarding pages of connected users), we took a biased sample with 10,000 users of each gender groups (men, women, unknown gender). We repeated the sampling procedure five times, to test for robustness to sampling error. We crawled the profile pages of all sampled users, and collected who they follow, and whom they are followed by. Gender of followers and followed users were identifies with the same approach outlined above.

**Measures and estimation**

The main variables of interest in our article is the gendered pattern of behavior, which we operationalize as the probability of being female given behavior. Several studies had adopted a similar approach of using an empirical typicality measure as an explanatory variable, in a wide range of empirical problems, from the phonological typicality of words (36) to the typicality of music (37), careers (38), businesses (39), or restaurants (40). Typicality has been used to investigate gender as well (27, 41). We selected variables that capture the most relevant aspects of behavior in open source software development. We use variables that represent choices reasonably under the control of the individual.

For the measurement of femaleness of behavior we included groups of variables describing professional ties, level of activity, and areas of specialization. We included the following variables describing professional ties: the number of collaborators and followed persons, separately for three gender categories: females, males, and unknown gender. We included variables describing general levels of activity: the number of pushes, the number of pull requests opened, the number of own repositories, and the number of repositories of others where the individual contributed anything (number of touched repositories). To capture the specialization of activity, we used principal component analysis of programming languages, where variables represented the number of times a given programming language was used by the individual. We identified six principal components, representing six typical specializations that software developers can concentrate in. These were robust in five different samples. See SI and S3 specifically.

We used the Random Forest regression tree classifier to predict the probability that an individual is female, as a function of his/her collaboration, activity, and specialization variables. See SI and S4 for procedures. We label this variable "femaleness" for short. The advantage of this method over more conventional classifiers (such as logistic regression or latent variable discriminant analysis) is that the classifier is based on decision trees, and not linear models. This is analogous to the difference between a model that takes all interactions of variables into account over a model that enters only first order main effects. The Random Forest classification was moderately accurate – behavior in open source is not drastically different by gender. The area under the ROC curve was 0.71, which was consistent across five samplings, and decreased to no less than 0.67 with 5% and 10% swapped gender. Variable importance scores were also robust to gender classification error. See S5 and S6. This is a moderate classification performance, which is weaker than classic instruments devised to measure gendered behavior (25) (AUC for inkblots test = 0.94, for combined test = 0.96), but similar to the performance of gender classifiers based on internet messaging (28) (AUC = 0.72), graphic design works (27)(AUC = 0.72), or biometric gender prediction based on screen swiping (29) (AUC = 0.71).

The most important behavioral aspect for femaleness prediction is gender homophily: the number of female collaborators. This variable has both the highest variable importance and the highest odds ratio. With one standard deviation increase in the number of female collaborators, the odds of being female increases by 1.84 (p=0.000). Other gender-coded collaboration tie variables are far less important, corroborating findings of others that female homophily is a marked phenomenon in fields where women are underrepresented (19). Specializations of programming languages are important components of gendered behavior, although contradicting stereotypical assumptions. Front-end specialization (work on the look of interfaces) is assumed to be feminine, while back-end (work on algorithms and data procedures under the hood) is considered to be more male. We identified two principal components of each specialization, and found that there is one pair of front-end and one back-end specialty that is more male, while there is another pair of front-end and back end specialty that is more female. For variable importance and odds ratios see S7.

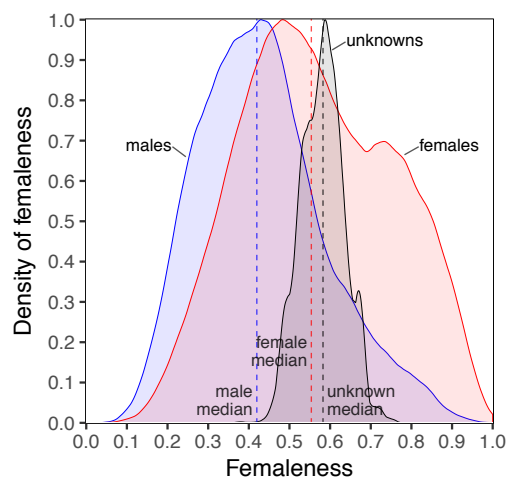

**Fig. 1.** Density of femaleness by gender categories.

Males have a median femaleness of 0.42, females 0.55, and the highest is unknown gender, with a median femaleness of 0.58. This indicates that users who do not reveal their gender are either females, or males with a decidedly female-like behavioral profile. Users with unknown gender also show the narrowest range of femaleness (0.32 to 0.76; compared to males: 0.07 to 0.96; and females: 0.06 to 0.99).

To assess the validity of the femaleness measure we tested whether for users with unknown gender there is a relationship between femaleness of behavior and the presence of female followers. If the femaleness of behavior is visible and meaningful for interacting users in open source programming, then we expect gender homophily to operate along the intensity of behavioral femaleness as well, not only manifest categorical gender. The correlation between femaleness and the proportion of female followers is weak, but positive significant (R=0.086, Spearman rho=0.037). The correlation is stronger using observations with ten or more followers (R=0.177, Spearman rho=0.147). Taking a logit model predicting the presence of female followers by femaleness, at the minimum of observed femaleness the probability of having any female followers is 0.22 (95% CI: 0.16-0.28), while at the maximum it is 0.34 (95% CI: 0.29-0.39). Using a negative binomial prediction, we expect 0.33 female followers at the minimum of femaleness (95% CI: 0.21-0.46), and 1.15 at the maximum (95% CI: 0.87-1.44).

To test the argument that one dimension is inadequate to capture varieties of gendered behavior (42, 43), we also used a decision tree prediction approach. With this approach we identified 16 classes of typical gendered constellations of behavioral variables. See SI, and especially S8.

Our dependent variables are success and survival. Our success measure is the total number of times other users have starred (bookmarked as useful) repositories of our focal user, during the entire career. A star is a statement of usefulness: interest from another user to easily locate and to utilize the given repository in the future. Since success and our behavioral variables co-evolve during the career, causal arguments cannot be tested. We measured survival by re-visiting all users' pages exactly one year after the end of our data collection, and recording the number of actions taken by the user over this one year. If a user did not make any actions on the site for one year, we recorded exit for that user; otherwise we marked the user as survivor. Users seldomly close their accounts (0.3% of users), since keeping an account is free. In the case of survival we can test causal hypotheses, as behavior precedes cessation.

Our measure of success is an over-dispersed count variable, thus we use a negative binomial model specification. Moreover, we also know that many users of GitHub are not interested in accumulating stars for repositories, but use the platform for other purposes (e.g. as a personal archive); in other words users are a mixture of two latent classes: one interested in achieving success, and one without such interest. We therefore estimated a zero-inflated negative binomial model, where we separately modeled excess zeros with a logit model, and the accumulation of stars with a negative binomial model. We also tested the robustness of our findings with an OLS model with the log of success as the dependent variable.

We estimate the following equation for our zero-inflated negative binomial mixture model:

$$\begin{cases} P(Y_i = 0) = \pi_i + (1 - \pi_i) \cdot (1 + k\lambda_i)^{-\frac{1}{k}} \\ P(Y_i = n) = (1 - \pi_i) \cdot \Gamma\left(Y_i + \frac{1}{k}\right)(k\lambda_i)^{Y_i} \Big/ \Gamma\left(\frac{1}{k}\right) \Gamma(Y_i + 1)\Gamma(1 + k\lambda_i)^{Y_i + \frac{1}{k}} \end{cases}$$

,

where $Y_i$ is the number of stars accumulated by user $i$ for own repositories, $\Gamma$ is the gamma distribution, $k$ is a dispersion parameter, and $n$ is a natural number $> 0$. We can model $\pi_i$ and $\lambda_i$ as functions of independent variables. For $\pi_i$ - the model for the zero component - we specify the following logistic regression with a logit link function:

$$logit(\pi_i) = \gamma_0 + \gamma_g x_{gi} + \gamma_b x_{bi} + \gamma_{gb}(x_{gi} x_{bi}) + \gamma_n x_{ni} + \gamma_{gn}(x_{gi} x_{ni}) + \gamma_c x_{ci}$$

and for the count model we use an identical specification:

$$log(\lambda_i) = \beta_0 + \beta_g x_{gi} + \beta_b x_{bi} + \beta_{gb}(x_{gi} x_{bi}) + \beta_n x_{ni} + \beta_{gn}(x_{gi} x_{ni}) + \beta_c x_{ci}$$

where $x_g$ is the female gender category (for women $x_g$=1, for men $x_g$=0), and $x_b$ is the femaleness of behavior from our random forest prediction. As an auxiliary test for the presence of discrimination by categorical gender, we added a variable that records the relative frequency of the first name of the user (relative to the total number of users of the same gender) – an approach recently taken to measure discrimination in patenting (44). If discrimination is by categorical gender, we expect women to be significantly disadvantaged in proportion to the frequency (easy recognizability) of their names. We expect that women with names like "Mary" (the most common female name) are more disadvantaged than women with names like "Maddie" (one of the least common female names). We thus include $x_n$ as the normalized logged relative frequency of first name within gender: $x_{ngi} = log\left(\frac{f_i}{N_g}\right) / \max(x_n)$, where $f_i$ is the overall frequency of the first name of user $i$, and $N_g$ is the overall number of users of gender $g$.

Finally, $x_{ci}$ stands for control variables. Our control variables represent alternative explanations connecting gender and outcomes: Tenure (number of years since joining) might favor men, as women tend to have shorter tenure (and drop out). The level of activity (number of own repositories and number of repositories where the user contributed) might also favor men, as women usually have less time to devote to professional activities. Social ties (number of followers and collaborators) might also favor men, as gender homophily is expected. Finally, we measure the total number of potential bookmarkers as the number of developers who worked with the same programming languages as our focal subject. A developer with a large potential audience might gather stars more easily for his or her repositories.

To test the adequacy of a single continuous dimension to represent behavioral femaleness, as opposed to a multi-categorical measurement, we identify multiple classes of femaleness with a decision tree prediction approach. We then include a set of binary indicator variables representing decision tree classes, with the most gender-balanced class being the reference category.

We estimate a logit model for survival with an identical specification to the success model :

$$ln\frac{P(Y_i = 1|X)}{1 - P(Y_i = 1|X)} = \beta_0 + \beta_g x_{gi} + \beta_b x_{bi} + \beta_{gb}(x_{gi} x_{bi}) + \beta_n x_{ni} + \beta_{gn}(x_{gi} x_{ni}) + \beta_c x_{ci}$$

where $Y_i = 1$ for users with sustained activity over one year after data collection, and $Y_i = 0$ for cessation. The independent variables are defined in the same way as described above.

**Results**

Considering gender as a category (females and males) for success, women on average received 8.76 stars, and men received 13.26, however, this difference is not statistically significant, neither by an F-test (F=2.208), nor by a bivariate zero inflated negative binomial model entering only an intercept and gender category (female=1, male=0) in both the zero inflation model (gender coefficient z= 0.488), and the count model (gender coefficient z= 0.835). Women, however, have a statistically significant disadvantage in the probability of survival: 92.8% men survived one year after our data collection, while only 88.2% of women (odds ratio=0.575, Chi-squared=126.1).

The femaleness of the pattern of behavior is significantly negatively related to success, using both a t-test (t=-5.337), and a zero-inflated negative binomial model (zero inflation model z=23.947; count model z=-12.365). Femaleness is also negatively related to survival (bivariate logit model z=-9.875)

Turning to multivariate models, Fig. 2 shows point estimates of expected success and expected probability of survival for gender-related variables from five model specifications. All variables are measured on the 0-1 scale, making estimates comparable. In our full models – ZINB models for success (S9) and logit models for survival (S11) – the coefficient for being female shows no consistent relationship with outcomes. In our main models of success and survival (model 1 with variables shown on Fig. 2 and additional control variables), females are not significantly disadvantaged compared to males. In fact, our success model shows a weak positive coefficient (0.62, p=0.049). We tested the robustness of this finding by adding binary indicator variables for decision tree classes representing typical gendered behavioral patterns (model 2), or adding all programming language use frequencies (model 3). We also re-estimated model 1 (both for success and survival) with randomly swapped genders. We estimate model 4 by using the same variables as in model 1, but randomly swapping the gender for 5% of developers in the sample with known gender, and in model 5 swapping 10% gender. Both model 4 and model 5 report 95% confidence intervals from 100 trials. Of the five models, only models 4 and 5 (with 5% and 10% randomly swapped gender) show significant disadvantage for females in survival. Our findings for success were robust with an OLS specification predicting log(success+1) as well (S10).

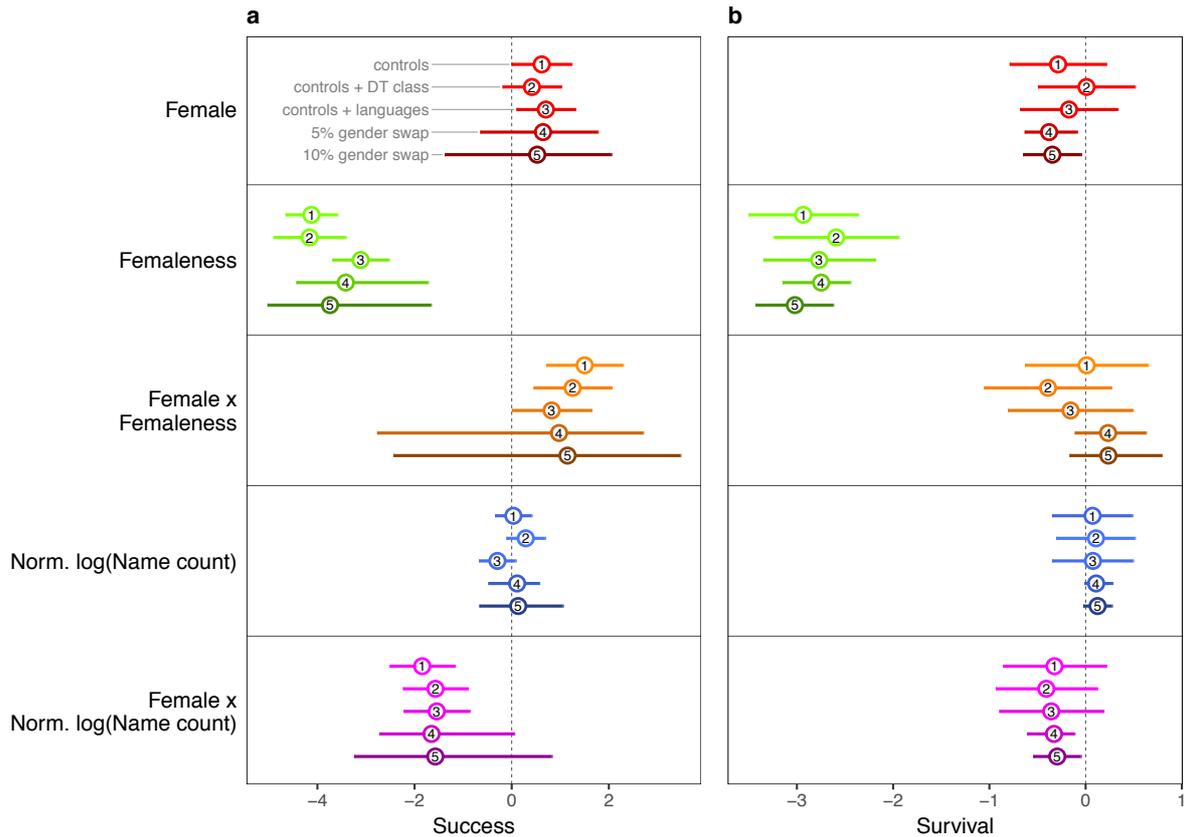

**Fig. 2.** Point estimates, with 95% CIs, for variables related to gender (variables are listed on the vertical axis). Panel a. shows coefficients from count models of zero-inflated negative binomial models predicting success (the number of stars received), while panel b. shows log odds ratios from logit models predicting survival over a one year period following our data collection. Labels of five specifications (identical for success and survival models) are shown in the legend. The first model enters gender variables and controls, the second enters controls and categorical gender behavior classes from the decision tree analysis, the third enters controls and 23 variables recording programming language use. The fourth is identical to the first, but with data with 5% gender swaps, and the fifth is with 10% gender swaps. For the fourth and fifth models confidence intervals show the 2.5 – 97.5 inter-quantile range from 100 simulated datasets.

While categorical gender is not a consistently significant predictor of outcomes, the femaleness of behavior is in all models for both success and survival. Femaleness of behavior is a strong negative predictor of both success and survival, and it is the only coefficient related to gender that is consistently and significantly different from zero. Fig. 3 shows predictions for success and survival along the range of femaleness, keeping all other variables constant at their means. The difference between females (red line) and males (blue line) is small compared to the difference along the range of femaleness.

First, consider success at the median for both males and females (Fig. 3 panel a). Taking the predicted success of males at the median is 2.53 (stars for their repositories), for females the prediction at their median femaleness is 1.07. Taking the male prediction as 100%, the expected success of females is 42.3% of that. The disadvantage is 57.7% points, of which 8.9% points are due to categorical gender, and 48.8% points are due to difference in femaleness. In other words, only 15.4% of the expected female disadvantage in success is due to categorical gender, and 84.5% is due to femaleness of behavior. Considering the same decomposition for probability of survival (Fig. 3 panel c), we see a smaller disadvantage for women: 6.1% points, of which 4.0% points is doe to categorical gender, and 2.1% due to differences in femaleness (34.8% of the expected disadvantage in survival).

Males are also disadvantaged by their gendered behavior. Considering the interquartile range of femaleness, the expected success of males at the first quartile of femaleness (0.32) is 4.16 stars, while the same expectation at the third quartile (0.52) is only 1.51 stars, which is 63.7% less. For females the predicted success at the first quartile of femaleness (0.43) is 1.84 stars, while at the third quartile (0.72) it is only 0.51 stars – a difference of 72.2%. For survival the same inter-quartile disadvantage for males is 2.7%, for females it is 8.8%.

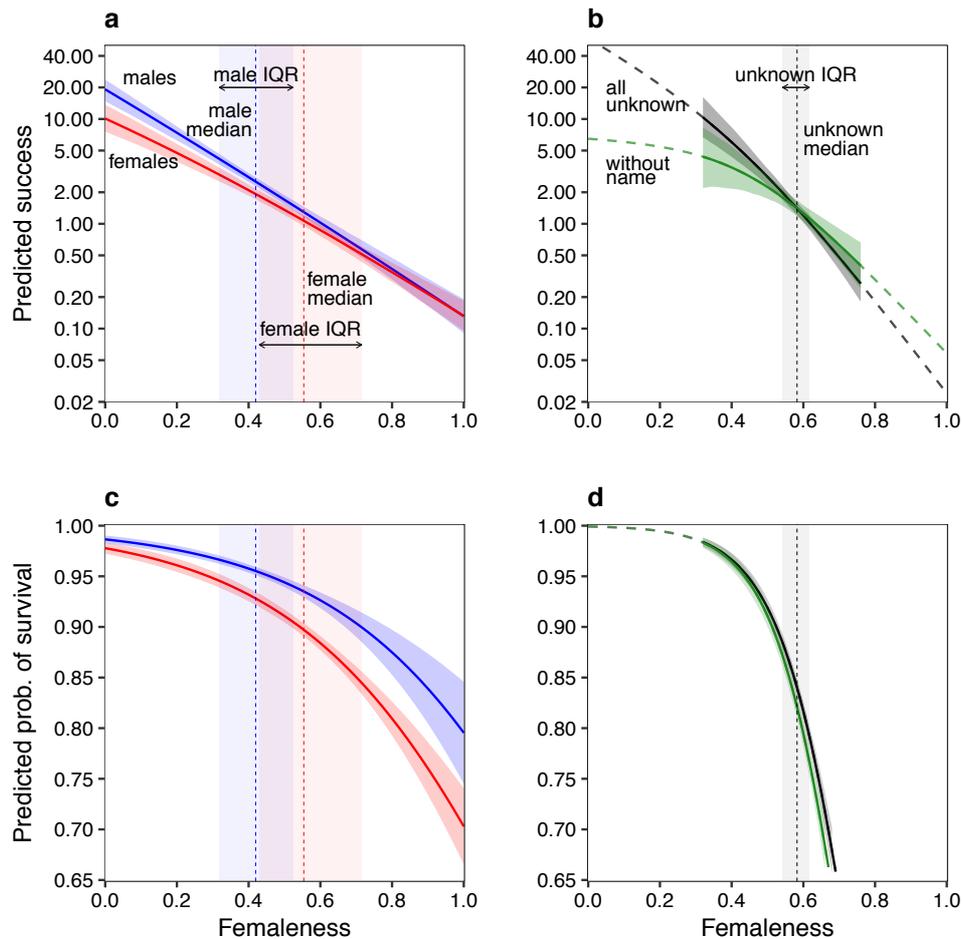

**Fig. 3.** Marginal predictions for femaleness by gender category from model 1 from Fig. 2. of success and survival, with fixing all other variables at their means. Panels a. and c. uses data for males and females, panels b. and d. uses data of users with unknown gender. Prediction is only shown for the observed range of femaleness. Vertical dashed lines indicate medians of femaleness, and shaded vertical bars show the interquartile range (IQR).

The coefficient of the interaction between female gender and femaleness is positive for success, but not significantly different from zero for survival (considering model 1). This indicates that the penalty for femaleness is higher for males overall than for females. (The female disadvantage over the interquartile range is nevertheless higher than males because of the wider spread of femaleness for females.)

Using the frequency of first name shows some evidence of discrimination in success, but not in survival. The interaction of being female and having a frequent name is negative, while the coefficient for name frequency itself is not significant, indicating that it is only women, who suffer a disadvantage if their name is more common, and thus their gender is easier to recognize. The prediction for a woman with the rarest name is 2.74 stars, while the prediction for a woman with the commonest name is only 0.95 stars – a 65.5% lower success.

Fig. 3 also shows predicted outcomes for users with unknown gender. To predict outcomes for unknowns, we use a specification identical to model 1, without variables for categorical gender and name frequency (see S12 and S14). Again, our findings about success were robust with an OLS specification predicting log(success+1) (see S 13). As apparent on Fig. 3 panel b and d, the femaleness disadvantage is also demonstrable for those who do not reveal their gender. At the first quartile of femaleness (0.54) the expected number of stars is 1.99, while at the third quartile (0.62) it is only 1.03 stars – a 48.0% drop. The disadvantage for survival is even more severe: a reduction of 10.4% across the interquartile range (compared to 2.7% for males, and 8.8% for females). These results are robust if we restrict our analysis to those users who do not reveal any name, and omit those who do reveal a name that was not listed in the US baby name dataset.

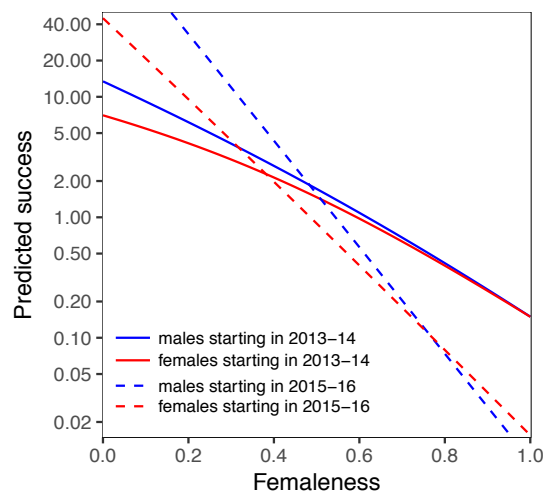

**Fig. 4.** Marginal predictions from zero inflated negative binomial model (model1) of success, for femaleness by gender category, separately for those who started in 2013-14, and those who started in 2015-16.

As a simple analysis of a time trend, we introduced a variable capturing those who started in the years of 2015 and 2016 (as opposed to starting in 2013 or 2014), and entered interactions for this time variable with categorical and behavioral gender into our model of success (S15). The resulting marginal predictions are shown of Fig. 4. We do not see evidence for a mitigating trend in the effect of behavioral gender, in fact, it seems that inequalities in success along the behavioral gender dimension have become more severe.

**Discussion**

Our study reveals that disadvantage in open source software development is a function of gendered behavior. We found consistent negative coefficients for femaleness, and only weak support for categorical discrimination. Femaleness of behavior is not only a disadvantage for women: men and users with unidentifiable gender are just as disadvantaged along this dimension. This is an important finding, as thus far the relative importance of categorical and behavioral gender have not been studied in the context of software development.

Our findings have important consequences for policy and interventions in gender inequalities in software development. First, in the short term, attempts to set quotas for women in software companies will not address the component of inequality that is related to gendered behavior. Increased proportion of women eventually might lead to the flattening of the slope of the relationship between behavioral femaleness and outcomes. A higher proportion of women can lead to questioning stereotypes, more visible female success stories in conventionally male types of behavior, and decisions to re-classify types of work that are now packaged in masculine-feminine stereotyped specialties.

Second, we should re-think the place of coding schools for women that are becoming widespread. These schools are typically training women in specialties that already have a number of women working in them, and thus might perpetuate the disadvantage of women by their femaleness of behavior (45). Another component of these schools is that they contribute to gender homophily by creating more women-to-women ties.

Third, users, and especially women, should re-think the benefits of hiding their gender identity online. It seems that the inequalities stemming from gendered behavior impact those just as much who hide their gender identity. A hidden gender identity can prevent discrimination by categorical gender, but it might also lead to a lack of trust and exclusion from projects, that might be behind the higher exit rate of such users.

Finally, it is important to distinguish gendered behavior from gendered free choice. We were composing our measure of gendered behavior out of variables that could be controlled by the individual, but we don't want to leave the impression that these traits are fully under the control of the individual. It is likely that the reasons behind the high (and increasing) negative slope of femaleness of behavior is due to constrained choice and deep-rooted stereotypes, rather than free choice.

**Acknowledgements**

This research was supported by the "Intellectual Themes Initiative" of Central European University, 2016-18. We thank Michael Szell for his assistance in collecting the data, and we thank participants of seminars at the Department of Network and Data Science at CEU, at the Institute for Analytical Sociology at Linköping University, and at the Institute for Social Research and Policy at Columbia University.